\documentclass[twocolumn,english,reprint]{revtex4-1}
\usepackage[T1]{fontenc}
\usepackage[utf8]{inputenc} 
\usepackage{color}
\usepackage{amsmath}
\usepackage{amsfonts}
\usepackage{graphicx}
\usepackage{physics}
\usepackage{lineno} 
\usepackage{babel}
\usepackage{babelbib}
\usepackage{bm}

\makeatletter

\newcommand{\UJvar}{U_{J}}
\newcommand{\en}{\varepsilon}
\newcommand{\vk}{\bm{k}}
\newcommand{\vhn}{\bm{\hat{n}}}
\newcommand{\vng}{\bm{n}_g}
\newcommand{\vngo}{\bm{n}_{g0}}
\newcommand{\vhp}{\bm{\hat{\varphi}}}
\newcommand{\vpm}{\bm{\varphi}_m}
\newcommand{\vecone}{\bm{1}}
\newcommand{\I}{\mathcal{I}}
\newcommand{\T}{\mathcal{T}}
\newcommand{\phioff}{\gamma_{ij}(\vpm)}

\makeatother

\begin{document}

\title{Weyl Josephson Circuits}

\author{Valla Fatemi}
\email{valla.fatemi@yale.edu}
\affiliation{Department of Applied Physics, Yale University, New Haven, CT 06520, USA}
\author{Anton R. Akhmerov}
\affiliation{Kavli Institute of Nanoscience, Delft University of Technology, P.O. Box 4056, 2600 GA Delft, The Netherlands}
\author{Landry Bretheau}
\affiliation{Laboratoire des Solides Irradi\'es, Ecole Polytechnique, CNRS, CEA/DRF/IRAMIS, Institut Polytechnique de Paris, F-91128 Palaiseau, France}

\begin{abstract}

We introduce Weyl Josephson circuits: small Josephson junction circuits that simulate Weyl band structures.
We first formulate a general approach to design circuits that are analogous to Bloch Hamiltonians of a desired dimensionality and symmetry class.
We then construct and analyze a six-junction device that produces a 3D Weyl Hamiltonian with broken inversion symmetry and in which topological phase transitions can be triggered \emph{in situ}.
We argue that currently available superconducting circuit technology allows experiments that probe topological properties inaccessible in condensed matter systems.

\end{abstract}

\maketitle

\section{Introduction}

Topological classification is a building block in our understanding of condensed matter systems~\cite{kane_quantum_2005,hasan_colloquium:_2010,qi_topological_2011}.
Electronic matter that may exhibit topologically non-trivial ground states include insulators, semimetals, and superconductors.
These ideas have been rapidly introduced to many other physical systems, such as quantum circuits~\cite{leone_topological_2008,roushan_observation_2014, nash_topological_2015} and metamaterials~\cite{lu_topological_2016}.
We now understand that materials are just one platform in which to explore the physics and potential applications of topologically non-trivial systems.

Topologically protected degeneracies are particular points of interest.
Three dimensional Weyl bands are an example in which pairs of nodes in the spectrum (e.g. degeneracies of two bands) with opposite topological charge persist over an extended region of Hamiltonian parameter space: a full band gap may open only when nodes of opposite charge have merged.
When a material exhibits these bands it is called a Weyl semimetal~\cite{wan_topological_2011,armitage_weyl_2018}. 
The peculiar topology of Weyl semimetals results in a host of unusual physical observables, including surface dispersion arcs, strong Berry curvature effects, and responses under inter-band excitation~\cite{yu_determining_2016,chan_photocurrents_2017,de_juan_quantized_2017,ma_direct_2017}.
A menagerie of connected topological band structures with nodal manifolds protected by symmetry have since been proposed and investigated~\cite{lin_spin-1_2015,bradlyn_beyond_2016,chang_nexus_2017}.

However, known Weyl semimetal materials often have additional physics that can obscure phenomena associated with the Weyl nodes (see Ref. ~\cite{bernevig_recent_2018} for a discussion), and as a result these topological Hamiltonians are being sought out on different platforms~\cite{lu_experimental_2015,rocklin_mechanical_2016,tan_realizing_2017,zhang_double-weyl_2018,lu_probing_2019}.
In particular, superconductor-based devices and circuits have shown promise for realizing topological states~\cite{georgescu_quantum_2014}, including investigation of topological concepts more generally~\cite{leone_topological_2008,leone_merging_2013,roushan_observation_2014,tan_realizing_2017,tan_topological_2018} and for the protection of quantum information~\cite{kitaev_unpaired_2001,ioffe_topologically_2002,kitaev_protected_2006,gladchenko_superconducting_2009}.
Recently, high-transmission multi-terminal Josephson junctions were proposed to realize Weyl Hamiltonians~\cite{van_heck_single_2014,yokoyama_singularities_2015,riwar_multi-terminal_2016,meyer_nontrivial_2017,klees_microwave_2020}, but these mesoscopic devices require control of microscopic electronic states.
Circuits based instead on linear elements~\cite{lu_experimental_2015,imhof_topolectrical-circuit_2018,zhao_topological_2018} and standard Josephson tunnel junctions are well-developed and offer great flexibility for designing \emph{in situ} tunable Hamiltonians~\cite{vool_introduction_2017}.

Here, we describe an approach to construct small Josephson junction circuits that simulate single-particle Hamiltonians with designable dimensionality and \emph{in situ} controllable symmetry classes.
We then describe specific applications that simulate Weyl band structures that can be tuned through a topological phase transition. 
We thus dub these Weyl Josephson circuits.
The proposed circuits exhibit all the features of Weyl band structures: protected energy degeneracies, divergent Berry curvature near those degeneracies, and the quantized topological invariant.
Finally, we argue the topologically non-trivial nature of the circuits can be measured in experiments that are unavailable to real materials.

\section{Building the Circuits} \label{sec_build}

\subsection{General Circuit Considerations}

The circuits we consider contain Josephson tunnel junctions as nonlinear inductive elements as well as linear capacitances such as those associated to the tunnel junctions.
These circuits have the Hamiltonian~\cite{vool_introduction_2017}
\begin{align}
\begin{split}
\hat{\mathcal{H}} =& \frac{(2e)^2}{2}\left(\vhn-\vng\right)^{T}\mathcal{C}^{-1}\left(\vhn-\vng\right) \\
&-\sum_{i,j} E_{J_{ij}} \cos \left(\hat{\varphi}_i - \hat{\varphi}_j - \phioff \right)\label{generalham} 
\end{split}.
\end{align}
Here the capacitive part depends on the number operators $\hat{n}_i$ counting the Cooper pairs on each circuit node $i$.
The gate voltages offset each charge operator in the capacitive energy by an offset charge $n_{gi}$.
For compactness, we use vector notations $\vhn$, $\vng$, and an inverse capacitance matrix $\mathcal{C}^{-1}$ that encompasses details of the circuit.
The Josephson energy of each junction is $E_{J_{ij}}$, $\hat{\varphi}_i$ are the phase operators canonically conjugate to $\hat{n}_i$. 
The phase offsets $\gamma_{ij}(\bm{\varphi}_m)$, depend only on the magnetic fluxes $\bm{\varphi}_m$ up to a gauge choice. 
For convenience we measure fluxes in units of the superconducting flux quantum $\hbar/2e$.

Similar to a Bloch Hamiltonian $\mathcal{H}(\vk)$, the Hamiltonian $\mathcal{H}(\vng,\vpm)$ of Eq.~\eqref{generalham} is periodic in the continuous offset variables, $\vpm$ and $\vng$ (for $\vng$ this is true up to an integer translation in $\vhn$). 
Regardless of the device geometry, $\mathcal{H}(\vng,\vpm)$ satisfies the following symmetry constraints.
First, it has a charge inversion symmetry $\I \mathcal{H}(\vng,\vpm) \I^{-1} = \mathcal{H}(-\vng,-\vpm)$, with the inversion operator $\I=\delta_{n,-n'}$ in the charge basis, and $\delta_{n,m}$ the Kronecker delta.
Additionally, the time-reversal symmetry $\T$ reads $\T \mathcal{H}(\vng,\vpm) \T^{-1} = \mathcal{H}(\vng,-\vpm)$, with the antiunitary operator $\T$ being complex conjugation in the charge basis.
Circuits with equal elements may have other unitary symmetries~\cite{ivanov_interference_2001} that we leave for later work.

We utilize these symmetry relations to set the symmetry properties of our simulated Hamiltonian.
A selection of offset parameters comprise effective crystal momenta $\vk$ that span the Brillouin zone of the simulated Hamiltonian, with the remainder being control parameters. 
To emulate a time-reversal symmetric dispersion relation we choose to vary all magnetic fluxes, while keeping the offset charges constant.
Unless $\vng\in \left\{0,1/2\right\}$, this results in a dispersion relation that lacks inversion symmetry.
A dual way to realize a time-reversal symmetric dispersion relation is to vary $\vng$ while keeping $\vpm$ constant.
Varying all magnetic fluxes and offset charges at once corresponds to an inversion-symmetric dispersion relation that lacks time reversal symmetry. 
In App.~\ref{Ap_TRS} we demonstrate that a standard flux qubit or a Cooper pair box realizes a minimal time-reversal symmetry breaking Weyl dispersion relation.
Finally, fixing a mixed set of fluxes and offset charges at a value not equal to $0$ or $1/2$ results in a fully asymmetric dispersion relation.

\subsection{Constructing a Time-Reversal Symmetric Weyl Josephson Circuit}

\begin{figure}[h]
\begin{centering}
\includegraphics[width=0.7\columnwidth]{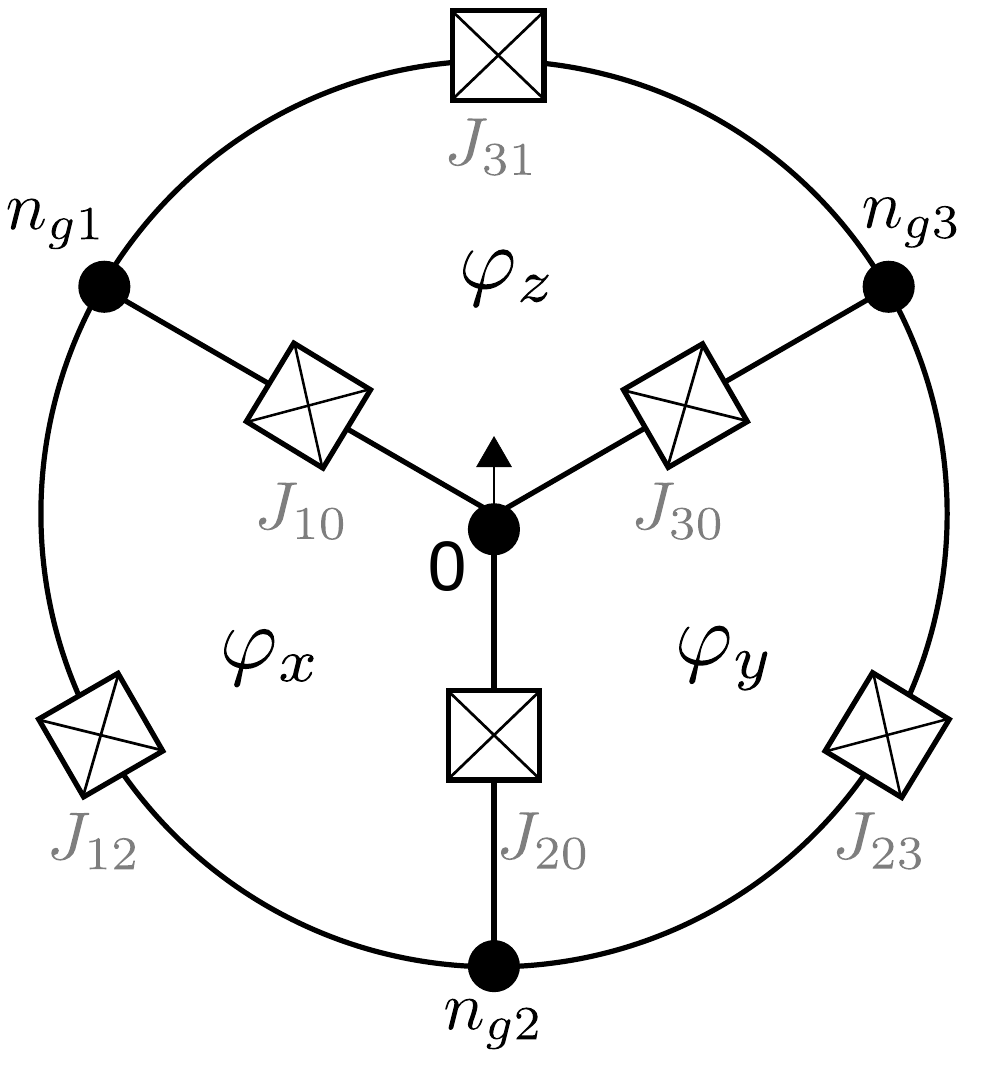}
\par\end{centering}
\caption{ \label{fig:ISBcircuit}
Diagram of a Weyl Josephson circuit that simulates an inversion-symmetry-breaking Weyl semimetal.
Six Josephson tunnel junctions $J_{ij}$ (which include both a capacitance and a Josephson energy) connect nodes $i$ and $j$, where $i=0$ is the reference node.
The active nodes $i\in\{1,2,3\}$ are labeled with their gate charge parameter $n_{gi}$, and the three loops are threaded by the reduced magnetic fluxes $\vpm=(\varphi_{x},\varphi_{y},\varphi_{z})$, expressed in units of the reduced flux quantum $\hbar/2e$. 
}
\end{figure}

According to the symmetry properties of Josephson circuits, a dispersion relation with Weyl points and a time-reversal symmetry may manifest only in a circuit with at least 3 independent  magnetic fluxes or 3 offset charges.
For convenience, we choose a circuit that satisfies both these requirements in a symmetric way, shown in Fig.~\ref{fig:ISBcircuit}~\footnote{We note that this circuit was previously inspected~\cite{feigelman_superconducting_2004} in a strongly symmetry-dependent context (e.g. all Josephson and charging energies must be equal) in order to generate a quadratically protected degeneracy at the point where the circuit is tuned so that six charge states are equal in energy. We are unaware of prior work pointing out the existence of Weyl nodes in this circuit. We briefly discuss the quadratic degeneracies of~\cite{feigelman_superconducting_2004} in Appendix~\ref{ap_nexus} }.
We now choose $\vk=\vpm=(\varphi_{x},\varphi_{y},\varphi_{z}) \in \left[0,2\pi\right]^3$ and utilize $\vng = (n_{g1},n_{g2},n_{g3}) \in \left[0,1\right]^3$ as control parameters.

We derive the Hamiltonian of this circuit in App.~\ref{ap_ham}.
To obtain its spectrum and eigenstates, we truncate the Hilbert space to contain several lowest energy charge states, and numerically diagonalize the Hamiltonian projected on this subspace.
To present the generic physics more transparently we take all Josephson energies as well as capacitive energies to be equal and consider the charge-dominated regime, $E_{C}\gg E_{J}$, with $E_C=(2e)^2/2C$ the charging energy of a single junction. 
In this situation, the minimal configuration of interest is in the vicinity of four charge states tuned to similar energy via the gate charges $\vng$. 
For example, choosing the globally uncharged state $\left|n_1 n_2 n_3\right\rangle=\left|000\right\rangle $ (all active nodes have zero net charge) and all the singly charged states $\left\{ \left|100\right\rangle ,\left|010\right\rangle ,\left|001\right\rangle \right\} $ (a single Cooper pair on any one of the three islands), we obtain the following simplified Hamiltonian
\begin{widetext}
\begin{equation} 
\hat{\mathcal{H}}(\vpm,\vng)=-\frac{E_{J}}{2}\left(\begin{array}{cccc}
-\lambda (n_{g1} + n_{g2} + n_{g3}) & 1 & 1 & 1\\
1 & -\lambda ( 1 - n_{g1}) & e^{i\varphi_{x}} & e^{-i\varphi_{z}}\\
1 & e^{-i\varphi_{x}} &  -\lambda ( 1 - n_{g2}) & e^{i\varphi_{y}}\\
1 & e^{i\varphi_{z}} & e^{-i\varphi_{y}} & -\lambda ( 1 - n_{g3})
\end{array}\right),\label{eq:Hamiltonian4x4}
\end{equation}
\end{widetext}
where $\lambda  = E_C / E_J \gg 1$. 
All four charge states are approximately degenerate at $\vngo=\frac{1}{4}\vecone$, where $\bm{1}=(1,1,1)$ is the unit diagonal vector. Thus, when $\lambda\|\vng-\vngo\|<1$, flux bias has a significant effect on the system.
With this Hamiltonian in hand, we now turn to an inspection of its physics. 

\section{Topological Spectrum and Invariants}

\subsection{Energy spectrum}

\begin{figure*}[t]
\includegraphics[width=1.9\columnwidth]{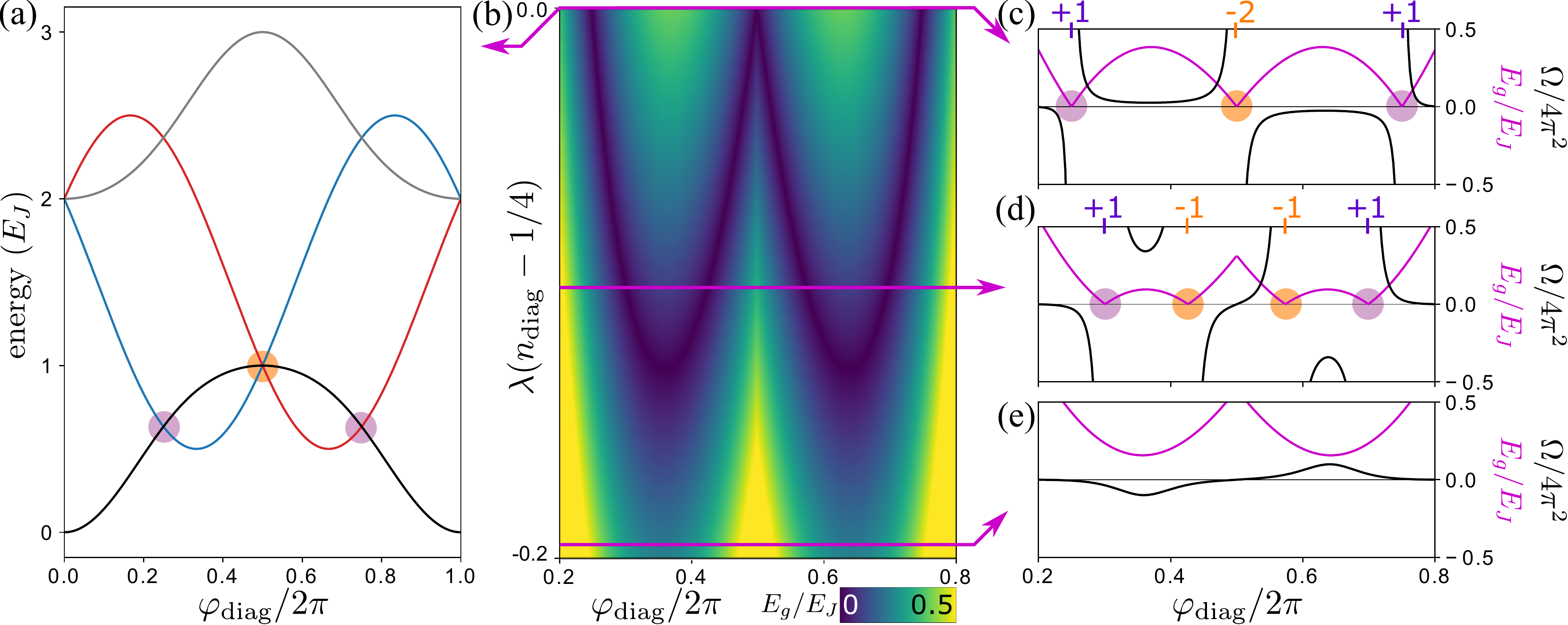}
\caption{\label{fig:simple_spectrum_and_node_motion}
(a) Energy spectrum in the charge-dominated regime from \eqref{eq:Hamiltonian4x4} at $\vng=\vngo$ as a function of flux along the main diagonal $\varphi_\textrm{diag}$. 
Line color is determined from wave-function continuity~\cite{akhmerov_connecting_2017}.
(b) Excitation gap from the ground state to the first excited state ($E_g$) as a function of symmetric gate charge $\vng-\vngo = n_\textrm{diag} \bm{1}$ and $\varphi_\textrm{diag}$. 
(c-e) Line-cuts of the excitation gap $E_g$ (pink) and the Berry curvature $\Omega$ (black) along $\varphi_\textrm{diag}$ for the indicated gate charge values.
}
\end{figure*}

We first inspect the dispersion relation of the Hamiltonian~\eqref{eq:Hamiltonian4x4}.
When the charge states are electrostatically degenerate ($\vng=\vngo$), the ground state is doubly degenerate at two points  $\vpm=\tfrac{\pi}{2} \bm{1},\ \tfrac{3\pi}{2}\bm{1}$, with $\bm{1}=(1,1,1)$, and triply degenerate at the point $\vpm=\pi \bm{1}$, as shown in Fig.~\ref{fig:simple_spectrum_and_node_motion}(a) \footnote{The triply degenerate point appears to realize an effective spin-1 system predicted in~\cite{bradlyn_beyond_2016}. We do not inspect it in detail due to its sensitivity to symmetry and the need to tune to a particular offset charge point.}.
These nodes lie along the major diagonal of the Brillouin zone $\vpm=\varphi_\textrm{diag}\bm{1}$ due to the spatial symmetry of the circuit.

Varying the gate charge induces a topological phase transition.
We consider uniform gating, $\vng=\vngo + n_\textrm{diag} \bm{1}$ with $0<n_\textrm{diag}<0.25$, which retains the symmetry that leaves all nodes on the major diagonal of the Brillouin zone.
As shown in Fig.~\ref{fig:simple_spectrum_and_node_motion}(b-e) the triply-degenerate point splits into two doubly-degenerate points while still lying along the major diagonal of flux. 
At a certain point, $n_\textrm{diag} - 1/4 \approx - 0.13 \lambda^{-1}$, the Weyl nodes converge and annihilate each other, signalling the topological phase transition. 
Beyond this point the ground state is gapped from the higher energy manifold for all flux configurations.

The Weyl nodes also survive asymmetric tuning of $\vng$ and non-uniformity of the Josephson energies.
This can, for example, induce the Weyl nodes to shift off of the major flux diagonal.
Figure~\ref{fig:nodes_and_chern}(a) shows these effects for a configuration with no particular symmetry except the time-reversal symmetry guaranteed by the choice of effective crystal momenta.
The Weyl node locations are indicated by the spheres (the meaning of the color, the topological charge, will be described in the next subsection). 
While offset charge detuning can generally induce the topological transition, the necessary amount of such detuning depends on the particular parameters of the device, particularly $E_J/E_C$ (see Appendix~\ref{ap_exp_transmon}).

\subsection{Quantum Geometry and Topological Invariants}

We now investigate topological aspects of the ground state wavefunction, which is accomplished by inspecting its Berry curvature $\bm{\Omega}$.
The Berry curvature relates the overlap between derivatives of the wave-function $\ket{\Psi}$ with respect to different Hamiltonian parameters, as in equation \eqref{eqnBerry3}:
\begin{align}
\bm \Omega = \Im \sum_{i,j,k
}
\bra{ \partial_{\varphi_{i}}\Psi} \ket{\partial_{\varphi_{j}}\Psi} \epsilon_{ijk} \bm e_k
\label{eqnBerry3} \\
C\left[S\left(\vpm\right) \right]      = \frac{1}{2\pi} \oint_{S} \bm{dS} \cdot \bm{\Omega} \label{eqnChern}
\end{align}
where $\epsilon_{ijk}$ is the Levi-Civita symbol and $\bm e_k$ is the unit vector in direction $k$.
Because of its closeness to a second-derivative of the wave-function, it is referred to as a curvature and described as a quantum geometry.
Its integral over a surface $S\left(\vpm\right)$ enclosing a Weyl point (equation \eqref{eqnChern}) is a Chern number $C\in \mathbb{Z}$ that determines the topological charge, or equivalently the chirality, of the Weyl point.
Nonzero Berry curvature and Chern number both are associated with physical observables which will be described in Section~\ref{sec_expts}.

The Berry curvature of the ground state is plotted alongside the energy gap in Figure~\ref{fig:simple_spectrum_and_node_motion}(c-e). 
In the topological phase (Fig.~\ref{fig:simple_spectrum_and_node_motion}(c,d)), the Berry curvature diverges where the ground and excited states are degenerate. 
Integrating the Berry curvature around each node (equation~\ref{eqnChern}), we determine the topological charges indicated in panels (c,d).
The topological phase transition occurs when nodes of opposite charge converge and annihilate each other, as can be seen in the transition from~\ref{fig:simple_spectrum_and_node_motion}(d) to~\ref{fig:simple_spectrum_and_node_motion}(e).

There are four distinct Weyl points in most of the topologically nontrivial situations described here, which we calculated to have integer Chern numbers indicated in Figure~\ref{fig:simple_spectrum_and_node_motion}(c-d).
Doubly-charged points may also exist, such as in Fig.~\ref{fig:simple_spectrum_and_node_motion}(a,c).
Crucially, like-charged nodes exist at opposite crystal momenta $C\left[S\left(\vpm\right)\right] = C\left[S\left(-\vpm\right)\right]$. 
This observation remains true in circuits with nonuniform Josephson elements, such as that shown in Figure~\ref{fig:nodes_and_chern}(a), where the position and charge of the four nodes are indicated by colored spheres.
This establishes that this circuit is a simulator of a broken-inversion-symmetry Weyl semimetal with preserved time-reversal symmetry~\cite{belopolski_minimal_2016}, as expected from the symmetry-based design.

\subsection{Summary of Extensions}

The appendices describe additional analysis which we summarize here. 
In Appendix~\ref{ap_exp_transmon}, we go beyond the charge-dominated regime of the circuit.
A useful finding from this is that the volume of offset charge space hosting a topological phase is maximized when $E_J \sim E_C/2$.
As well, the topological phase is naturally robust to the experimental degree of disorder in the junctions (Appendix~\ref{ap_exp_disorder}). 
We further discuss  circuits that simulate Weyl band structures with broken time-reversal symmetry (Appendix~\ref{Ap_TRS}) and more exotic semimetals that depend on additional symmetries (Appendix~\ref{ap_exotics}) or exist in higher dimensions (Appendix~\ref{ap_4D}).

\section{Experimental Path Forward} \label{sec_expts}

In the previous sections, we outlined the principles of construction, the resulting spectrum, and abstract topological aspects of a Weyl Josephson circuit. 
In this section we describe three experimental observables to probe the topologically non-trivial character: transition spectra, adiabatic responses measuring Chern number, and non-adiabatic responses measuring Berry curvature.
The latter two both rely on the consequences of Berry curvature on the system response function~\cite{gritsev_dynamical_2012}. 
In regards to circuit parameters and noise robustness, all these experiments are accessible using modern nanofabrication and measurement techniques (see Appendix~\ref{ap_experimental} for details). 
In Appendix~\ref{Ap_TRS} we also discuss the application of these experiments to time-reversal symmetry breaking Weyl circuits, which would be attractive for initial investigations due to their simplicity. 
Some of these experiments are not possible to perform on a real material, highlighting the complementary nature of parametric simulation with superconducting quantum circuits.

\subsection{Microwave Spectroscopy of Topological Phase Transitions}
The simplest experiment is microwave spectroscopy of the circuit under otherwise static conditions.
The goal is to measure the energy spectrum as a function of $\vk$ and detect degenerate points.
The smoking gun is to observe that these Weyl nodes survive when varying the gate charges over a finite range and finally annihilate following a topological phase transition. 
In practice, one needs to perform such spectroscopy within $\sim [0,50]$ GHz (the exact range will depend on the particular circuit parameters chosen).
One possibility is to perform the standard two-tone spectroscopy used in circuit QED by coupling the circuit to a superconducting resonator~\cite{wallraff_strong_2004,kubo_strong_2010,janvier_coherent_2015,tosi_spin-orbit_2019}.
Another approach is Josephson spectroscopy, which uses a voltage-biased Josephson tunnel junction as an on-chip microwave spectrometer~\cite{edstam_josephson_1994,leppakangas_tunneling_2006,billangeon_very_2007,bretheau_exciting_2013,bretheau_supercurrent_2013}.
These two methods are complementary since Josephson spectroscopy is better suited for high frequencies (typically within $[2,100]$ GHz) while two-tone spectroscopy works better at lower frequencies (typically within $[0.1,30]$ GHz).

\begin{figure*}
\includegraphics[width=1.9\columnwidth]{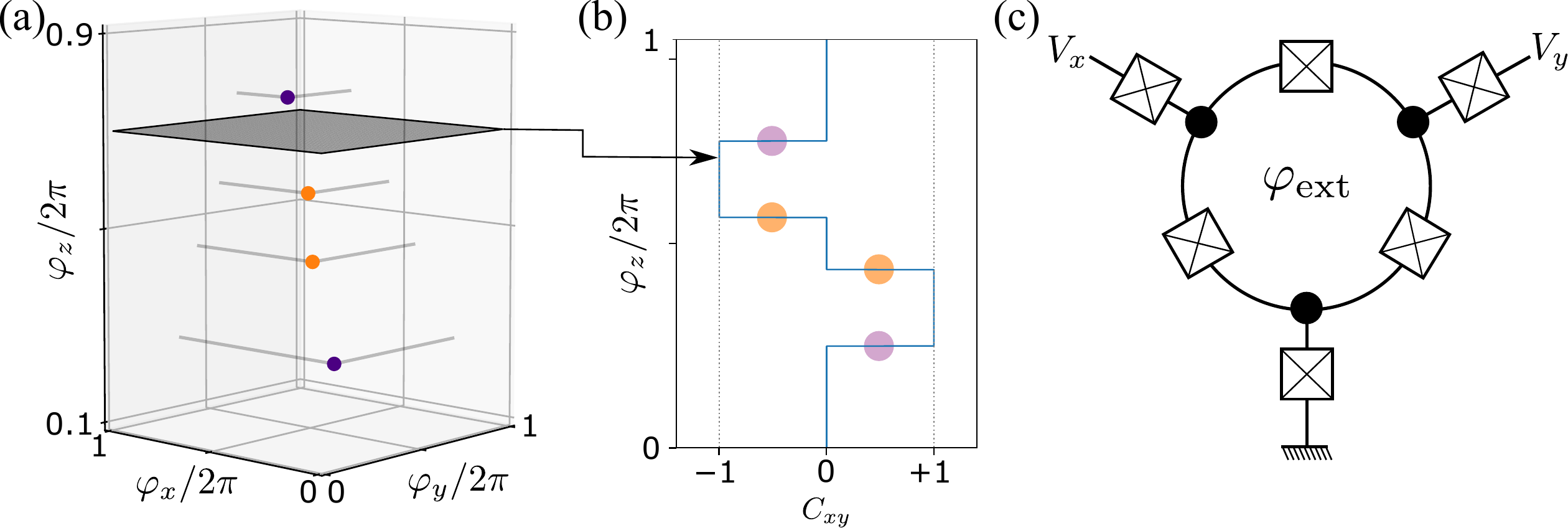}
\caption{\label{fig:nodes_and_chern}
(a)
Location of the Weyl points in flux-space for $\lambda n_\textrm{diag} = 0.08$ and $E_{J_{12}} = 1.7 E_J$. 
Like-charged nodes exist at opposite momenta, as in an inversion-symmetry broken Weyl semimetal. 
The grey plane indicates the integration area that determines the Chern number $C_{xy}$ for $\varphi_{z}=0.7\times2\pi$.
(b) 
Chern number as a function of $\varphi_{z}$, $C_{xy} (\varphi_{z})$, for the same parameters as in (a).
(c) 
Broken-open circuit for transconductance measurements.
The upper leads are biased by dc voltages $V_x$ and $V_y$ referenced to the grounded lower lead, and the loop is threaded by a flux $\varphi_{ext}$.
Measurement of quantized dc transconductance would reveal the Chern number~\cite{riwar_multi-terminal_2016}. 
}
\end{figure*}

\subsection{Transconductance to Measure Chern Number}

A direct measurement of topological invariants is a stronger indication of a topologically nontrivial state.
The Chern number may be accessed by measuring the response due to adiabatic variation of Hamiltonian parameters, as we propose here with transconductance measurements.
The periodicity of the Hamiltonian means 2d planes in flux space bounded by the effective Brillouin zone constitute closed manifolds and therefore have a Chern number. 
Figure~\ref{fig:nodes_and_chern}(b) shows the Chern number $C_{xy} (\varphi_{z})$ for the $(\varphi_x,\varphi_y)$ plane as a function of $\varphi_z$ (now treated as a control knob), for the circuit in a topological phase. 
Whenever a Weyl node is crossed by tuning $\varphi_z$ (see Fig. 3(a)), the Chern number changes by an integer amount corresponding to the charge of that node. 
After a topological phase transition into a trivial phase, the Chern number is zero for any value of $\varphi_z$. 
These Chern numbers can be probed by performing transconductance measurements, as proposed by Riwar and collaborators for the case of multi-terminal high-transmission Josephson junctions~\cite{riwar_multi-terminal_2016}. 
Similarly, here one may break open two of the loops of the Weyl circuit and apply dc voltages (see Fig. 3(c)).
Using the ac Josephson effect, two of the fluxes can be varied adiabatically and linearly with time while keeping the third flux static.
If the rate of change of these two fluxes is incommensurate then one can sample a complete 2d plane within the Brillouin zone. 
This will lead to a dc current that is directly proportional to the Chern number, and therefore to a quantized transconductance~\cite{riwar_multi-terminal_2016,eriksson_topological_2017} without the complications of a nearby continuum~\cite{meyer_nontrivial_2017,repin_topological_2019}.
We note that this transconductance has a close relationship with Cooper pair pumps~\cite{geerligs_single_1991,leone_topological_2008,peryuchat_conductance_nodate} and may be useful for metrological applications~\cite{kaneko_review_2016,peryuchat_conductance_nodate}.

\subsection{Direct Berry Curvature Measurements}

Going further, the Berry curvature can be directly measured as a function of the quasi-momenta. 
Then, by simple integration, one could extract the Weyl nodes' topological charge. 
This approach is based on a theoretical breakthrough describing how Berry curvature can be observed from the nonadiabatic response of physical observables to the rate of change of an external parameter~\cite{gritsev_dynamical_2012,kolodrubetz_geometry_2017}.
This method was recently implemented in the field of quantum circuits using a basic system, a driven qubit, to observe topological phase transitions~\cite{roushan_observation_2014,schroer_measuring_2014}.
Therefore, our proposed experiment is feasible albeit more involved than the ones discussed before.
Indeed, a direct Berry curvature measurement involves coherent manipulation of the circuit's quantum states (superposition of ground and excited states) and therefore requires that the circuit exhibits sufficiently long coherence times ($\gtrsim 1~\mu$s).
A related approach is to use carefully designed absorption spectroscopy measurements relying on the same underlying physics~\cite{klees_microwave_2020}.

\section{Concluding remarks}

In this manuscript we have described a proposal for Weyl Josephson circuits: small Josephson tunnel junction circuits that simulate Weyl band structures, including \emph{in situ} triggerable topological phase transitions.
We have also described several experiments that probe the topological nature of the circuit.
All the necessary ingredients for an experimental implementation are in reach with modern nanofabrication and experimental techniques. 

This work leaves open questions on how far these ideas can be developed.
An immediate step is the classification of the available symmetry classes, including point group symmetries beyond inversion, to explore more exotic topological band structures such as the two described in Appendix~\ref{ap_exotics}. 
An important question to this end is whether we can create robust analogies to spins and spin-orbit coupling in order to create analogues to quantum spin Hall insulators~\cite{kane_z2_2005}.
As well, can small Josephson circuits simulate topological boundaries and their unique surface dispersions?
Finally, applying strong driving or dissipation, a standard tool in superconducting circuits, extends the system to symmetry classes that are hard to access in condensed matter systems~\cite{mirrahimi_dynamically_2014,leghtas_confining_2015}. 
The interplay of Floquet physics with topological ground states~\cite{giovannini_floquet_2019} and circuit Hamiltonians~\cite{jooya_graph-theoretical_2016} are active areas of study that can be combined in circuits like the one presented here.

These circuits exhibit clear parallels with theoretical proposals based on Andreev levels in a scattering region contacted by multiple superconducting leads~\cite{yokoyama_singularities_2015,riwar_multi-terminal_2016,meyer_nontrivial_2017,eriksson_topological_2017,repin_topological_2019,klees_microwave_2020}. 
It will be interesting to inspect whether such proposals generically have a suitable Josephson circuit analogue, particularly when restricting to the even-parity sector and without spin-orbit coupling.

We are convinced that Weyl Josephson circuits offer a versatile, tunable, and complementary platform to probe the physics of topologically non-trivial systems.
We hope that this work will stimulate mutually beneficial contact between the fields of superconducting quantum circuits and topological condensed matter physics.

\subsection*{Acknowledgements}
We thank L. Peyruchat, J. Griesmar, and \c{C}.~\"{O}. Girit for fruitful discussions and for sharing their related forthcoming work ~\cite{peryuchat_conductance_nodate}.
We also acknowledge helpful discussions with M. Devoret, N. Frattini, L. Glazman, P. Kurilovich, V. Kurilovich, K. Serniak, C. Smith, and U. Vool;  L. Shi and J. Song; M. Houzet and J. Meyer. 
We thank A. Eickbusch for assistance regarding the topological volume fraction calculations.
LB acknowledges support of Agence Nationale de la Recherche through grant ANR-18-CE47-0012 (JCJC QIPHSC).
Code and data available on Zenodo~\cite{valla_fatemi_weyl_2020}.

\subsection*{Author Contributions}
VF conceived the idea and developed the theory with guidance from AA and LB. VF, AA, and LB all wrote the manuscript.

\bibliographystyle{apsrev4-1}

\onecolumngrid

\appendix

\section{Six-Junction Circuit Hamiltonian} \label{ap_ham}

\subsection{Derivation}  \label{ap_ham_derivation}

Here we derive the Hamiltonian of the circuit in Figure~\ref{fig:ISBcircuit} via network analysis in the usual way by starting in the nodal flux basis~\cite{vool_introduction_2017}.
Using the reduced flux-quantum $\phi_0=\hbar/2e$ to rescale into a phase basis, the Lagrangian is $\mathcal{L}=\mathcal{L}_{C}-U_J$, where the Josephson part can be written 
\begin{align}
\begin{split}
-U_J\big(\bm \varphi , \vpm \big)= &  E_{J_{10}}\cos\varphi_{1}+E_{J_{20}}\cos\varphi_{2}+E_{J_{30}}\cos\varphi_{3} +E_{J_{12}}\cos\left(\varphi_{1}-\varphi_{2}-\varphi_{x}\right) \\
 &  +E_{J_{23}}\cos\left(\varphi_{2}-\varphi_{3}-\varphi_{y}\right) +E_{J_{31}}\cos\left(\varphi_{3}-\varphi_{1}-\varphi_{z}\right)
\end{split}
\end{align}
with $\bm{\varphi}=\left(\varphi_{1},\varphi_{2},\varphi_{3}\right)$ and $\vpm=\left(\varphi_{x},\varphi_{y},\varphi_{z}\right)$. 
The charging part of the Lagrangian reads
\begin{align}
\begin{split}
\mathcal{L}_C = & \frac{\phi_0^2}{2} \Big( C_{10}\dot{\varphi}_{1}^{2}+C_{20} \dot{\varphi}_{2}^{2}+C_{30} \dot{\varphi}_{3}^{2} +C_{12} \left(\dot{\varphi}_{1}-\dot{\varphi}_{2}\right)^{2}  +C_{23} \left(\dot{\varphi}_{2}-\dot{\varphi}_{3}\right)^{2} +C_{31} \left(\dot{\varphi}_{3}-\dot{\varphi}_{1}\right)^{2} \Big) \\
 = & \frac{\phi_0^2}{2}\dot{\bm{\varphi}}^{T}\mathcal{C}\dot{\bm{\varphi}}
\end{split}
\end{align}
where we introduce a convenient $3\times3$ capacitance matrix $\mathcal{C}$. 
For compactness we ignore constant terms in the Lagrangian and leave off the reference phase defined at the central node $\varphi_0$.

We next account for offset charge on each node (which may be controlled by capacitive gates that are formally defined here to have sufficiently small capacitance so as not to impact the circuit modes) and define canonically conjugate momenta in the usual way in order to come to a Hamiltonian:
\begin{align}
\hat{\mathcal{H}} & =\frac{4e^2}{2}\left(\vhn-\vng\right)^{T}\mathcal{C}^{-1}\left(\vhn-\vng\right)+\UJvar
\end{align}
where $\vhn=\left(\hat{n}_{1},\hat{n}_{2},\hat{n}_{3}\right)$, $\vng=\left(n_{g1},n_{g2},n_{g3}\right)$, and $\vhp=\left(\hat{\varphi}_1,\hat{\varphi}_{2},\hat{\varphi}_{3}\right)$.
Finally, it is convenient to choose a characteristic scale for the capacitances in order to write the charge term with a prefactor with units of energy. 
For simplicity, we choose the average junction capacitance $\Bar{C}$, which defines both the characteristic charging energy $E_C = (2e)^2/2\Bar{C}$ and the dimensionless inverse capacitance matrix $c^{-1} = \Bar{C}\mathcal{C}^{-1}$:
\begin{align}
\hat{\mathcal{H}} & =E_C\left(\vhn-\vng\right)^{T}c^{-1}\left(\vhn-\vng\right)+\UJvar
\end{align}

\subsection{Tight-Binding Model}  \label{ap_ham_tightbinding}

We now take advantage of the fact that the individual Josephson terms are equivalently represented as a sum of single-Cooper-pair translation operators: 
\begin{align}
E_{J_{ij}}\cos\left(\hat{\varphi_{i}}-\hat{\varphi_{j}}+\phioff\right)\ & = \frac{1}{2}E_{J_{ij}} e^{i\phioff}\ket{n_i,n_j+1}\bra{n_i+1,n_j} + h.c.
\end{align}
These become the hopping terms of the tight-binding model. 
In the charge-dominated regime, $E_{C}\gg E_{J_{ij}}$, the case of interest studied in the main text is in the vicinity of a four-fold charge degeneracy point, which is in principle analytically soluble.
After assuming all $E_{J_{ij}}=E_J$ we have the $4\times4$ Hamiltonian matrix in equation~\eqref{eq:Hamiltonian4x4}.
The characteristic polynomial for the eigenenergies $\en$ (in units of $E_J/2$ for convenience) of this matrix is
When more than four charge states must be considered, or outside the deep charging regime, the model must be solved numerically with more charge basis states. For this we employ the Kwant tight binding package~\cite{groth_kwant:_2014}.

\section{Experimental Considerations}  \label{ap_experimental}

\subsection{Beyond Deep Charging Limit}  \label{ap_exp_transmon}

In the main text, for illustrative purposes, we focus on a simplified Hamiltonian that provides a good approximation of the circuit in the deep charging regime ($E_J/E_C \ll 1$) near $\vng = \vngo = \frac{1}{4} \bm{1}$, with $\bm{1} = (1,1,1)$.
However, the regimes in which Josephson energies $E_J$ are comparable to or larger than the charging energies $E_C$ are important because they are easily accessible in experiment~\cite{orlando_superconducting_1999,koch_charge-insensitive_2007}. 
The limit $E_J\gg E_C$ additionally provides exponentially suppressed sensitivity to offset charge noise in simple circuits like the transmon~\cite{koch_charge-insensitive_2007}. 
We cannot strictly retain this feature in Weyl Josephson circuits due to the topological phase transitions that can be triggered by offset charge tuning.  

\begin{figure}
\includegraphics[width=0.95\columnwidth]{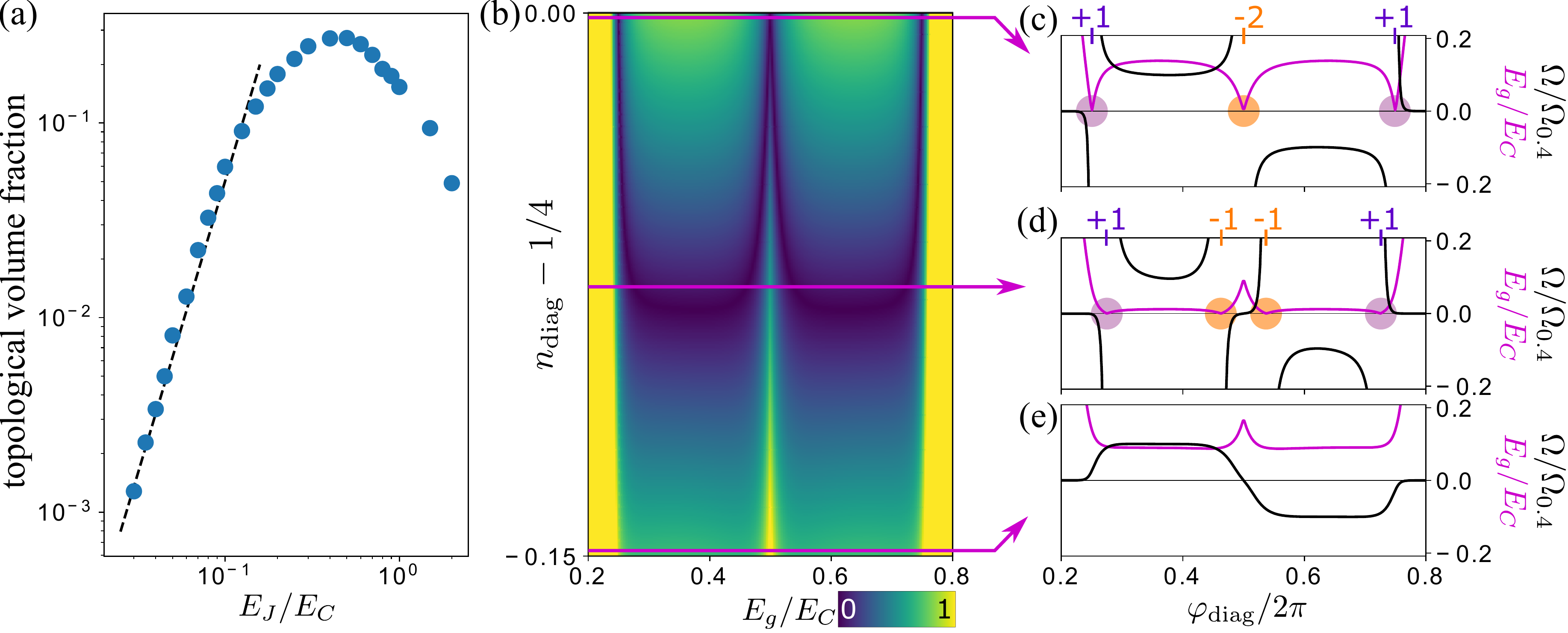}
\caption{\label{fig:Transmon}
(a) Topological volume fraction of the $\vng$ parameter space as a function of $E_J/E_C$.
This is estimated by finding ground-state degeneracies in $\vpm$-space on a 50x50x50 grid spanning the $\vng$ cube.
The dashed black line indicates a cubic trend. 
(b)  Excitation gap from the ground state to the first excited state ($\Delta E_g$) as a function of symmetric gate charge $\vng-\vngo = n_\textrm{diag} \bm{1}$ and $\varphi_\textrm{diag}$ for $E_J/E_C = 1.0$. 
Note that the characteristic scale is now set by $E_C$ rather than $E_J$.
(c-e) Line-cuts of $E_g$ (pink) and the Berry curvature $\Omega$ (black) along $\varphi_\textrm{diag}$ for the indicated gate charge values.
For presentation purposes the Berry curvature for each plot is normalized by $\Omega_{0.4}=10\left|\Omega\left(0.4\times2\pi\right)\right|$. 
}
\end{figure}

Nonetheless, choosing $E_J$ comparable to $E_C$ provides several experimental advantages for topological physics.
First, for $E_J \approx E_C/2$ the volume fraction of the $\vng$ parameter space in which the circuit exhibits ground state degeneracies in the Brillouin zone is maximized to about $0.27$ (Figure~\ref{fig:Transmon}(a)).
Maximizing this quantity is advantageous for experiments that must search for the topological regime by varying gate voltages on the islands. 
Second, the positions of the Weyl nodes become less sensitive to offset charge offsets (Figure~\ref{fig:Transmon}(b-e)), an advantage for experiments that may have moderate offset charge drift (see Appendix~\ref{ap_exp_noise} for more on this point).
Note that for $E_J \gg E_C$ the system is dominated by wells in the classical Josephson potential -- thus, degeneracy points indicate transitions to a new global minimum in which the lowest energy states of the two wells have no avoided crossing.
Third, the characteristic energy scale in the topological regime becomes sensitive to $E_C$, as indicated in Figures~\ref{fig:Transmon}(b-e).
This is convenient as $E_C$ can lowered by geometric circuit features independent of the junction's intrinsic capacitance.
All of these features make the moderate $E_J/E_C$ regime attractive for experiments. 
Curiously, as is visible in Figure~\ref{fig:Transmon}(b-e), the energy gap along contours connecting Weyl nodes varies with offset charge detuning despite the fact that the position of the nodes changes quite slowly. 
When this gap approaches zero, the nodes rapidly converge and annihilate.
We note that a previous work on Cooper pair pumps (equivalent circuit to Fig.~\ref{fig:ISBcircuit}(a)) investigated the effect of noise on pumping processes and found that $E_J\sim E_C$ was optimal for that circuit for different reasons than these~\cite{leone_cooper-pair_2008}.

\subsection{Junction Disorder}  \label{ap_exp_disorder}
Fabrication of Josephson tunnel junctions with aluminum electrodes and AlOx tunnel barriers is a relatively mature process. 
Junctions that are fabricated simultaneously and placed in the same region of a wafer can be made identical to within about a 2\% accuracy~\cite{kreikebaum_improving_2020}.
While the topological nature of the Weyl points implies a general robustness to variations in Hamiltonian parameters, it is reasonable to specifically ask whether a Weyl circuit is robust to the experimental degree of uncertainty. 
To this end, we model the six-junction circuit from the main text, targeting $E_J=E_C/2$ with all junctions identical.
To simulate disorder, we add random deviations in the junction area $A$, sampling from a Gaussian distribution with a conservative standard deviation of $10\%$. 
Note that errors in the junction area affect both $E_J \propto A$ and $E_C \propto A^{-1}$.
For each instance, we conduct a numerical search for spectral nodes in flux-space, fixing the gate charge value $\vng=\vngo$.
We find the topological phase in all 1000 tested instances signalling robustness to experimental fabrication uncertainty. 

\subsection{Noise Considerations}  \label{ap_exp_noise}

Both flux and charge noise are important to consider in any experimental proposal~\cite{krantz_quantum_2019}.
For each type, it is convenient to separately consider high-frequency ($\gtrsim kHz$) and low-frequency ($\lesssim Hz$, also known as "drift") parts.
High-frequency noise introduces measurable spectroscopic linewidth and reduces phase coherence. 
For the spectroscopy experiment proposed in Section~\ref{sec_expts}, linewidths limit the resolution with which any degeneracies can be determined. 
The third experiment (direct Berry curvature measurements) will be limited by the relationship between phase coherence time and the measurement time, the determination of which is beyond the scope of our work. 
Very high frequency noise, at the value of $E_g$, will introduce a finite excited state population. 
The rate of these unintended transitions sets a lower bound for the speed of the transconductance measurements.
In transmons, this rate has been lowered to the 100 $Hz$ range~\cite{serniak_direct_2019}, which is five to eight orders of magnitude below anticipated $E_g$ scales for the Weyl circuit.

Low-frequency noise, or drift, is a significant problem if it interferes with the typical time-scale of experimental scans. 
Flux drift is generally insignificant. 
Charge drift, however, can be large and has typical time-scales are of order tens of minutes or more for transmons~\cite{serniak_direct_2019,christensen_anomalous_2019}. 
For this reason, it may be sensible to choose $\vk=\vng$ rather than $\vpm$ for simulation of broken-inversion-symmetry systems, as it may be more convenient to fix the Hamiltonian control parameters for long periods of time. 
In either case, a solution will be needed if the necessary multidimensional measurement scans are slower than these timescales. 
This motivates the use of a cQED setup, which can take advantage of fast individual measurements to implement active feedback routines to correct for charge drift (see Ref.~\cite{shulman_suppressing_2014} for an example of active feedback improving spin qubit coherence).

\section{Additional Topological Josephson Circuits}
\subsection{Minimal Time-Reversal Symmetry Breaking Weyl circuits} \label{Ap_TRS}

\begin{figure}
\includegraphics[width=0.5\columnwidth]{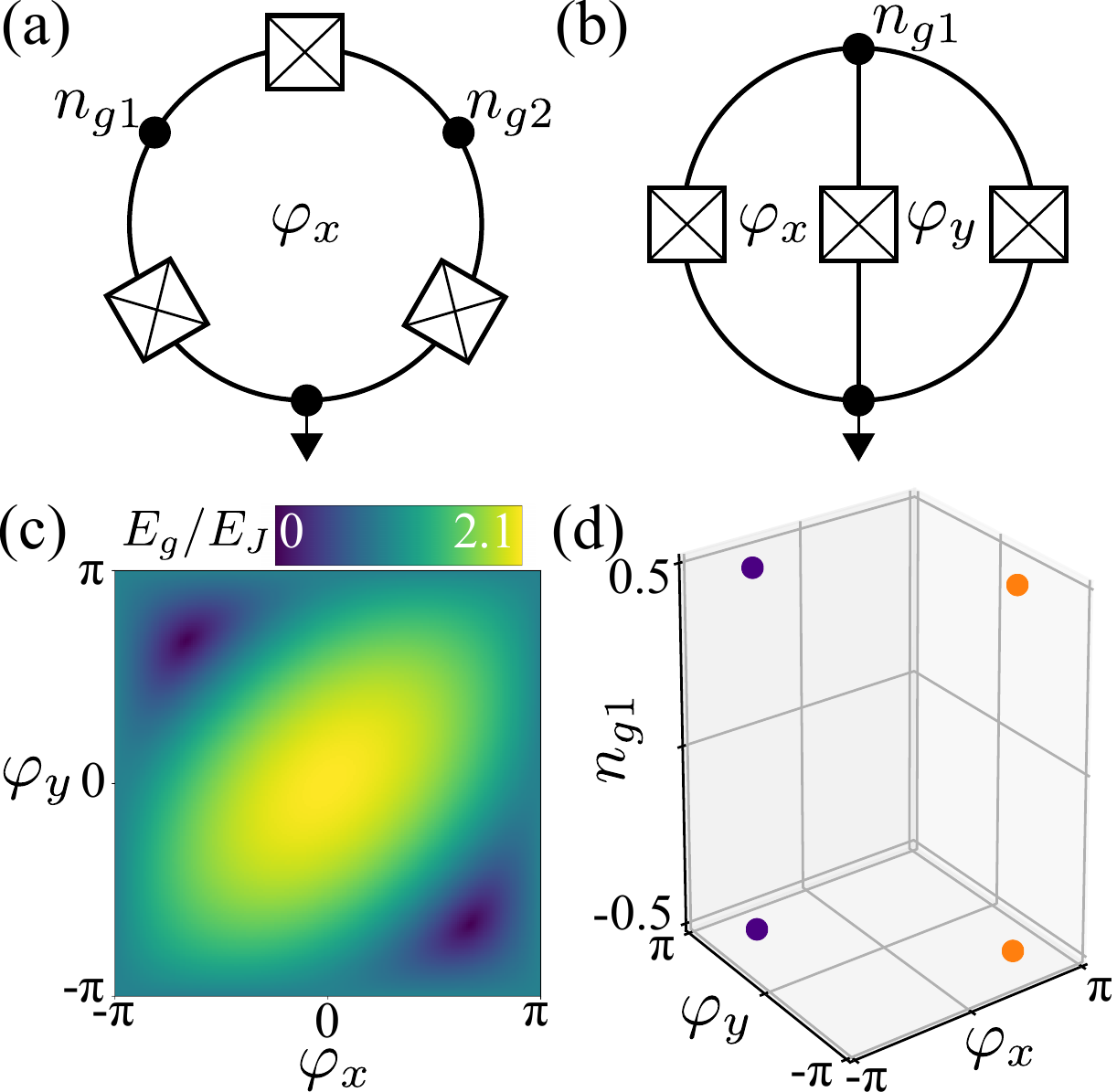}
\caption{\label{fig:Basic_Circuits}
(a) Flux qubit circuit diagram. 
(b) Gradiometric-SQUID Cooper pair box circuit diagram. 
(c) Excitation gap $E_g$ of the gradiometric-SQUID Cooper pair box as a function of the two fluxes for $n_{g1}=0.5$ and $E_J=E_C$ (all junctions identical).
(d) Location and charge (purple: +1, orange: -1) of the Weyl nodes in effective Brillouin zone for $E_J=E_C$. 
All junctions are identical. 
Note that the pair of nodes located at the top and bottom are in fact the same pair.
}
\end{figure}

A time-reversal breaking Weyl semimetal requires 3 offset variables, with both types present, to comprise $\vk$.
Minimal configurations satisfying this are drawn in Figure~\ref{fig:Basic_Circuits}(a,b). 
Both have three available offset parameters that can define a 3D quasimomentum space: the flux qubit has $(n_{g1},n_{g2},\varphi_x)$ and the gradiometric-SQUID Cooper pair box has $(n_{g1},\varphi_x,\varphi_y)$.
We focus on the gradiometric-SQUID Cooper pair box, with the flux qubit case having been investigated previously~\cite{leone_topological_2008,leone_topological_2008}.
In Figure~\ref{fig:Basic_Circuits}(c) we plot the energy gap as a function of the two fluxes for $n_{g1}=0.5$, and two nodes are clearly present. 
Using the same procedure as noted in the main text, we determine the topological charge of these nodes, which are shown in Figure~\ref{fig:Basic_Circuits}(d). 
Note that $n_{g1}=\pm0.5$ is the Brillouin zone boundary and so only two nodes are present, one of each charge.
Nodes of opposite charge are located at opposite quasimomenta, verifying that the circuit simulates the minimal broken-time-reversal-symmetry Weyl semimetal with preserved inversion symmetry~\cite{belopolski_minimal_2016}.

In these circuits, all available offset parameters are contained in $\vk$, so the Hamiltonian is otherwise fixed.
Thus, in order to induce a topological phase transition, one of the built-in circuit parameters, like a Josephson energy $E_J$, must be tuned.
For example in the gradiometric-SQUID Cooper pair box, if one of the Josephson energies is made larger than the sum of the other two a topologically trivial phase is found. 
One possible way is to use the electric field effect with a Josephson junction made from superconductor-semiconductor technology~\cite{shabani_two-dimensional_2016}.
Alternatively, one could replace a single Josephson element with a flux-tunable SQUID~\cite{leone_merging_2013} or gate-tunable Cooper pair transistor, whereby that element's offset charge tunes an effective Josephson energy. 
Consider 2d cuts of the band structure of the nodal line circuit of App.~\ref{ap_nodalline} as an example of this.
Note that inversion symmetry is not generally guaranteed in this situation.

Nonetheless, a topological phase transition is not necessary to observe physics in the topological regime, and the simplicity of these circuits is attractive for initial experiments. 
Indeed, the microwave spectroscopy and Berry curvature experiments proposed in Section~\ref{sec_expts} are relatively straightforward to consider. 
An equivalent to the transconductance experiment is less straightforward as one of the two linearly varying parameters must be $n_g$ (e.g. a dc current across a capacitor) in order to span a plane with a finite Chern number.

Finally, we note that two-island circuits, essentially equivalent to the one shown in Figure~\ref{fig:Basic_Circuits}(a), have a history of research regarding their topological degeneracies in the form of charge pumping. 
Charge pumping has been studied in both normal double-island devices~\cite{pothier_single-electron_1992,switkes_adiabatic_1999} and superconducting double-island devices~\cite{geerligs_single_1991,leone_topological_2008,leone_cooper-pair_2008,leone_merging_2013}.
Pumping in devices consisting of a single-island with multiple Josephson tunnel junctions (an extension of the circuit in Figure~\ref{fig:Basic_Circuits}(b)) have also been studied ~\cite{vartiainen_nanoampere_2007,mottonen_experimental_2008,gasparinetti_single_2012}. 
A similar experimental approach could also be employed in the circuits discussed in the main text, although we did not investigate this in detail.

\subsection{Higher-Symmetry Semimetals} \label{ap_exotics}

\subsubsection{Nodal Line Circuit}  \label{ap_nodalline}
\begin{figure}
\includegraphics[width=0.95\columnwidth]{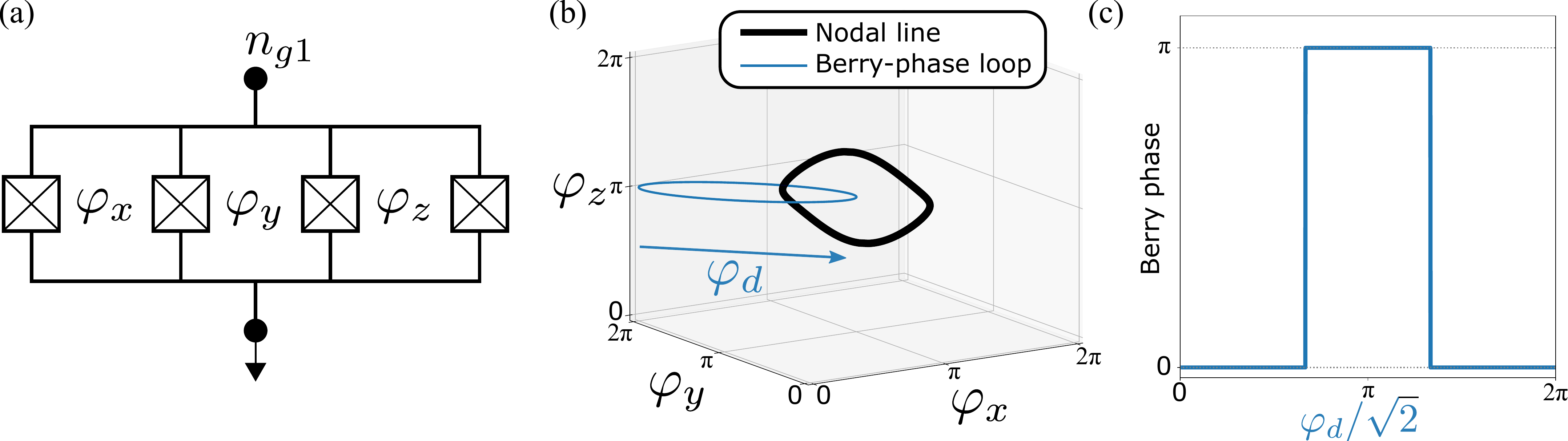}
\caption{\label{fig:nodal_loop_fig}
(a) Circuit that simulates a nodal line semimetal for $\vk=(\varphi_x,\varphi_y,\varphi_z)$ and $n_{g1}=0.5$.
(b) The closed-loop nodal line centered on $\vpm=(\pi,\pi,\pi)$ when one Josephson junction has double the Josephson energy than each of the other three.
Note we have shifted the Brillouin zone to more easily present the loop. 
In blue we depict an example loop through which we calculate the Berry phase with the variable long diameter $\varphi_d$. 
(c) Calculated Berry phase as a function of $\varphi_d$.
}
\end{figure}

In Figure~\ref{fig:nodal_loop_fig}(a) we show a circuit that simulates a nodal-line semimetal of the codimension type, requiring both inversion and time-reversal symmetry (see Ref. ~\cite{bernevig_recent_2018} for an overview and references to specific materials proposals and Ref.~\cite{leone_merging_2013} for discussion of a related circuit). 
Here, $\vk=(\varphi_x,\varphi_y,\varphi_z)$ while $n_{g1}$ is a control parameter, which mean the circuit generally satisfies the time-reversal symmetry condition. 
An inversion symmetry is present when $n_{g1}=0.5$, which we set.
In Figure~\ref{fig:nodal_loop_fig}(b) we show the line degeneracy between the ground and first excited states for the situation that three junctions are identical and the fourth has twice the Josephson energy.
It has the characteristic shape of a closed loop in momentum space.
The final requirement is a Berry phase of $\pi$ for a closed trajectory that links with the nodal loop. 
We check the Berry phase of a closed trajectory of variable diameter as indicated in the figure and indeed find a Berry phase of $\pi$ only when the trajectory links with the nodal loop (Fig.~\ref{fig:nodal_loop_fig}(c)).

\subsubsection{Possible Nexus Bands}  \label{ap_nexus}
\begin{figure}
\includegraphics[width=0.95\columnwidth]{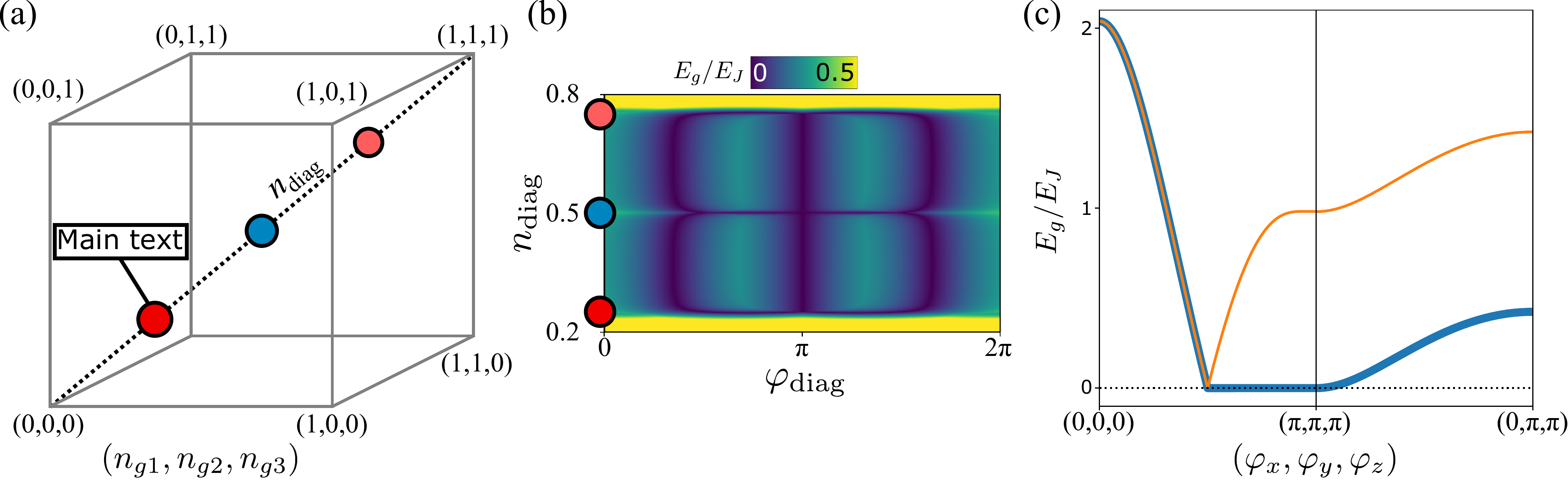}
\caption{\label{fig:Nexus}
(a) The 3D offset charge parameter space, with unique single-charge-state minima labeled on the corners. 
The vicinity of the red circle at $\vng = (1/4,1/4,1/4)$, is the subject of the main text.
The blue circle at $\vng = (1/2,1/2,1/2)$ is the position discussed in Appendix~\ref{ap_nexus}.
The vicinity of the pink circle at $\vng = (3/4,3/4,3/4)$, shares equivalent physics to the red circle but with a different set of charge states.
(b) The excitation gap as a function of both $n_\textrm{diag}$ and $\varphi_\textrm{diag}$, where we see that the two sets of Weyl spectra at the red and pink circles combine to form the nodal line phase at the blue circle.
Note that the structure of this plot shows the dual nature of gate charge and magnetic flux in this circuit. 
(c) Excitation gap $E_g$ in the deep charging regime for $\vng = (1/2,1/2,1/2)$.
The point $\vpm = (\pi,\pi,\pi)$ is the focus of Ref.~\cite{feigelman_superconducting_2004}. 
Detuning $\vpm$ from that point, we see that along some directions the ground state degeneracy is preserved (left) while in others it is broken quadratically (right).
}
\end{figure}

Returning to the six-junction circuit of the main text, we now describe a different high-symmetry offset charge point,  $\vng = \left( 0.5,0.5,0.5 \right)$, that exhibits an unusual spectrum. 
At this point, indicated by the blue circle in Fig.~\ref{fig:Nexus}(a), and in the deep charging regime, the basis states are \emph{six} singly- and doubly-charged states: $\left\{ \left|100\right\rangle ,\left|010\right\rangle ,\left|001\right\rangle \right\} ,\left\{ \left|011\right\rangle ,\left|101\right\rangle ,\left|110\right\rangle \right\}$.
Detuning $n_\textrm{diag}$ in either direction results in Weyl semimetal phases such as discussed in the main text, whose special charge degeneracy points are indicated by the red and pink circles in Fig.~\ref{fig:Nexus}(a,b). 
The three-state manifolds hosting Weyl nodes introduced at the red and pink circle are combined at the blue circle.
When all the Josephson junctions are identical, this combination produces open-ended line degeneracies in $\vpm$-space, as shown in Fig.~\ref{fig:Nexus}(c). 
This is quite reminiscent of a proposal for Nexus Fermions in materials~\cite{chang_nexus_2017}, which cites the need for certain point group symmetries.
Indeed, if all junctions are identical, there is a discrete three-fold rotational symmetry about the diagonal charge axis, several mirror symmetries, as well as an inversion center about the offset point $\vng= \left( 0.5,0.5,0.5 \right)$. 
We leave a deeper theoretical analysis of this situation for a future study.
Finally, we note that at a special effective momentum, $\vpm=\left(\pi,\pi,\pi \right)$ the ground-state degeneracy can be only be broken quadratically, which was the special subject of a previous paper by Feigel'man and collaborators proposing a dephasing-protected qubit~\cite{feigelman_superconducting_2004}.

\subsection{Higher Dimensions} \label{ap_4D}

\begin{figure}
\includegraphics[width=0.95\columnwidth]{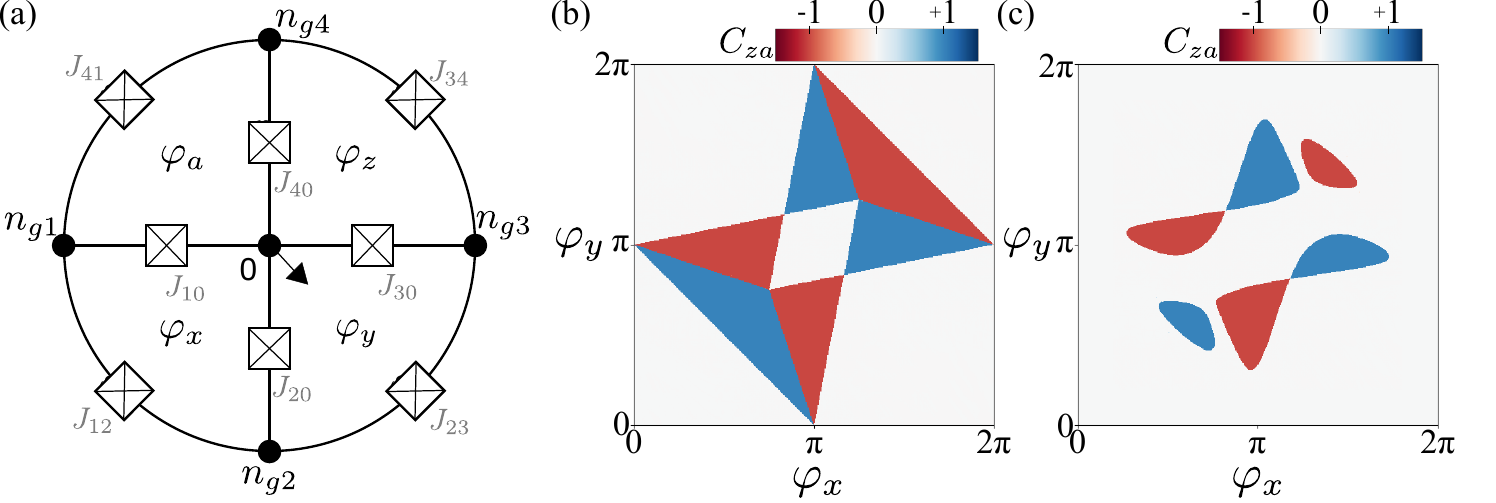}
\caption{\label{fig:higher_dim}
(a) Schematic for a circuit that simulates a four-dimensional Weyl system with broken inversion symmetry and preserved time-reversal symmetry.
(b-c) Chern number of the $(\varphi_{z},\varphi_a)$ plane as a function of the two remaining fluxes $\varphi_{x}$ and $\varphi_{y}$ and with the gate charges tuned to the 5-fold charge degeneracy point.
Panel (b) is the case for for a completely symmetric circuit, while panel (c) was generated with added randomness in the Josephson energies (sampled from a white distribution with maximum deviation of 15\%).
}
\end{figure}

The general approach to circuit construction for parametric simulation in Section~\ref{sec_build} is readily applied to Hamiltonians of higher effective dimensionality.
Here, we extend the symmetry class of the circuit in the main text -- that of preserved time-reversal symmetry and unprotected inversion symmetry -- to 4d. 
A similarly symmetric circuit with 4 nodes and 4 loops shown in Figure~\ref{fig:higher_dim}(a), where again we choose $\vk=\vpm$, accomplishes this.
This Hamiltonian exhibits 1d Weyl lines in the 4d Brillouin zone, in contrast to the 0d Weyl points found in the 3D case.
Moreover, we observe that Chern numbers defined in a 2d plane can be tuned as a function of two parametric fluxes, for which a pair of examples is shown in Figure~\ref{fig:higher_dim}(b-c). 
In this case, we tuned to a similar kind of charge-degeneracy point as in Eq.~\eqref{eq:Hamiltonian4x4}, now involving five charge basis states due to the fact that there are four islands: $\left\{ \left|0000\right\rangle, \left|1000\right\rangle ,\left|0100\right\rangle ,\left|0010\right\rangle ,\left|0001\right\rangle \right\}$.
Finally, we note that while this circuit is equivalent to the eight-junction Josephson ring modulator~\cite{schackert_practical_2013}, the studied cases have always been for extremely high $E_J/E_C$ so that the topological physics and $\vng$-dependencies are likely not practically observable.

\end{document}